# Robust Interaction Control of a Dielectric Elastomer Actuator with Variable Stiffness

Gianluca Rizzello, *Member, IEEE*, Francesco Ferrante, *Member, IEEE*,
David Naso, *Senior Member, IEEE*, Stefan Seelecke

*Abstract*— **This paper presents an interaction control algorithm for a dielectric elastomer membrane actuator. The proposed method permits to efficiently exploit the controllable stiffness of the material, and use it as a "programmable spring" in applications such as robotic manipulation or haptic devices. To achieve this goal, we propose a design algorithm based on robust control theory and Linear Matrix Inequalities. The resulting controller permits to arbitrarily shape the stiffness of the elastomer, while providing robust stability and performance with respect to model nonlinearities. A self-sensing displacement estimation algorithm allows to implement the method without the need of a deformation sensor, thus reducing cost and size of the system. The approach is validated on an experimental prototype consisting of an elastomer membrane pre-loaded with a bi-stable biasing spring.**

## I. Introduction

DIELECTRIC Elastomers (DEs) represent a class of active materials which have received significant attention in recent literature due to their large deformation (>100%), low power consumption (order of mW), high energy density, high bandwidth, and low cost [1]. A DE consists of a thin elastomeric film with compliant electrodes printed on its external surfaces. When a voltage is applied to the electrodes, the resulting electrostatic forces produce a change of the mechanical characteristics of the membrane, i.e., its stiffness and geometry [2]. Such a feature, together with the inherent compliance of the material, has made the adoption of DEs an interesting way to develop innovative devices with controllable stiffness in a variety of application fields, such as industrial robotics [3], soft robotics [4], [5], bio-inspired robotics [6], rehabilitation robotics [7], haptic systems [8],

The authors would like to acknowledge the support of Parker-Hannifin, BioCare Business Unit.

G. Rizzello and S. Seelecke are with the Department of Systems Engineering and the Department of Material Science and Engineering, Universität des Saarlandes, Saarbrücken 66123, Germany (e-mail: gianluca.rizzello@imsl.uni-saarland.de, stefan.seelecke@imsl.uni-saarland.de).

F. Ferrante is with the Computer Engineering Department, University of California Santa Cruz, 1156 High St., Santa Cruz, CA 95064 (e-mail: francesco.ferrante.2011@ieee.org).

D. Naso is with the Department of Electrical and Information Engineering, Politecnico di Bari, Bari 70125, Italy (e-mail: david.naso@poliba.it).

suspensions [9], and tunable resonators [10], just to mention a few. On the other hand, the use of DE material in real-life applications is currently limited by the high amount of voltage required for activation (several kV), sensitivity to environmental conditions, inconsistent resistance to fatigue, and strongly nonlinear and rate-dependent behaviors that limit their use in open loop. While the reduction of the activation voltage and the resistance to fatigue are being continuously addressed by advances in material science [11], [12], the compensation of uncertainties, nonlinearities, and external disturbances can be coped with feedback control strategies.

A large part of recent contributions on the subject focuses on DE position control [13]–[16], with advanced approaches ranging from open-loop control to bio-inspired methods. However, in many robotics applications, actuators operate in an unstructured environment in which unpredictable contact forces may arise, e.g., while grasping objects of different sizes and stiffness. In these cases, a position control scheme typically provides unsatisfactory performance, since it may generate unnecessarily high contact forces. Indeed, to improve the interaction between the actuator and the environment, force and position must be controlled simultaneously. This is the goal of Interaction Control (IC) strategies, which arbitrarily shape the dynamical relation between the system end-point variables such as position (or velocity) and force, rather than directly control one of the two [17]. This control paradigm has been successfully used in robotics (see, e.g., the survey papers [18], [19], or recent contributions [19], [20]). IC laws can be used to effectively compensate DE nonlinearities and dissipative phenomena, thus increasing speed and accuracy of controllable stiffness devices. In spite of these remarkable advantages, controllable stiffness devices based on DE are typically operated in open loop, and only few recent works have considered the application of IC to DE. One example is reference [21] which implements IC architecture for a ionic polymeric-metal composite artificial muscle, by using a PID control designed with standard tuning rules. Carpi *et al.* proposed in [7] a stiffness control strategy for a hand rehabilitation orthoses, by performing an online inversion of a black-box experimental representation of the transducer. The method is relatively simple to implement, but it does not take explicitly into account model nonlinearities, uncertainties, and dynamic effects such as viscoelasticity. Impedance control method for a DEA based on nonlinear design tools is proposed in [22], but its effectiveness is validated only by means of simulations.

Building on those recent papers, this manuscript presents a new approach to IC of DE based on robust control design tools. The approach is grounded on previous work by the authors on the design of position control systems for DEA with stability and performance guarantees [23]. The approach pursued in [23] is based on the reformulation of the DEA model as a Linear Parameter Varying (LPV) system, suitable for a class of design methods relying on efficient numerical optimization tools based on Linear Matrix Inequalities (LMI). It must be understood that the extension of our previous work on position control to the case of IC is not trivial. Indeed, when shifting the control goal from the position regulation to the shaping of the force-displacement characteristics, one needs to take into the account the effects of external forces acting on the system. In particular, relying on a partial state feedback controller, we first propose a set of Bilinear Matrix Inequality (BMI) conditions for the design of IC laws ensuring closed-loop stability, dynamic performance, and minimum control effort. Despite partial state feedback controllers are attractive for their relatively simple implementation and online computation effort, the design of such control laws via BMI represents a well-known NP-hard problem, thus unattractive from the numerical standpoint [24], [25]. Some approaches have been proposed in order to deal with this problem, including iterative LMI [26] or convexification [27], while in this paper we develop a novel method to convert the original BMI problem in a LMI eigenvalue problem, which can be efficiently solved via existing numerical tools. We point out that LPV theory has already been used for the solution of IC (e.g., [28]–[30]). However, the mixed performance-control gain tradeoff formulation proposed here has some important peculiarities. While standard approaches for IC of nonlinear systems rely on direct elimination of system nonlinearities [18], our approach is based on a linear law that requires much smaller implementation and online computation efforts. However, differently from other linear strategies, such as PID design based on linearized models, our approach provides guaranteed performance in the entire operating range.

Another distinctive feature of our approach is the use of a position self-sensing algorithm that, reconstructing DE position from voltage and current measurements, avoids the use of a position sensor to implement the IC. Being able to avoid the use of position sensors is highly appealing in practical applications involving force interaction, since typically adopted laser displacement sensors represent, in many cases, the most expensive part of the system. On the other hand, force measurements can be obtained in such settings via contact force transducers, e.g., a load cell, placed on the contact surface of the actuator. This represents a common solution adopted in IC [31]–[33]. The approach is extensively validated on an experimental platform specifically devised for the IC problem.

The reminder of this paper is organized as follows. Section II introduces the DEA used to validate the proposed theory. Section III deals with the design of the control algorithm, which is subsequently validated by means of simulations and experiments in Section IV. Finally, Section V discusses some conclusive remarks and possible future research directions.

## II. VARIABLE STIFFNESS DE

A DE can be viewed as a deformable capacitor with compliant electrodes. Depending on the specific application, DEs with different geometry and size can be used. In this paper, we employ an annular (or cone) membrane (see Fig. 1). Such a geometry has been successfully used in several applications and deeply analyzed in the literature [14]. A rigid passive frame hosts the membrane and makes it possible to achieve out-of-plane deformations, as depicted in Fig. 1(b) and (c) which show the undeformed and deformed states respectively. Fig. 2(b) illustrates the out-of-plane force-displacement characteristics of the membrane at steady state for minimum and maximum actuation voltages, namely 0 and 2.5 kV, emphasizing the relationship between voltage and stiffness. Voltages above 2.5 kV cannot be applied, since dielectric breakdown may occur.

In order to use the cone DE as an actuator, a mechanical biasing force needs to be applied on the inner disk. The biasing element has a significant effect on the overall actuator performance in terms of stroke and force. In this work we adopt a biasing mechanism based on a bi-stable spring, i.e., a nonlinear spring which admits two distinct stable equilibrium positions, since this design solution allows to significantly extend the achievable deformation [23]. A sketch of the actuator is shown in Fig. 2(c). The biasing spring is connected to the DE membrane via a spacer, by means of screw connections. When an external compressive force $f$ is applied, the DEA reacts with a blocking force $f_{DEA}$ given by

$$f = f_{DEA} = f_b - f_{DE}, \tag{1}$$

where $f_{DE}$ and $f_b$ are DE and biasing spring forces, respectively. A representation of the steady state external force in (1), for several DE voltages, is shown in Fig. 2(d). In this reference, the zero displacement corresponds to the situation in which the DE membrane is fully flat. The bi-stable spring makes the overall DEA characteristics in Fig. 2 significantly different from the ones in Fig. 2(b). Finally, Fig. 2(e) provides a photo of the overall system, in which it is possible to see the DE membrane, the screw connections used to apply voltage to the electrodes, the linear actuator used to apply an external force, and the load cell used as force sensor.

To implement IC, the voltage needs to be controlled in a way such that the DEA reacts to an external force similarly to a mechanical system with a desired force-displacement profile. To illustrate this concept, we assume that the desired characteristics for the DEA corresponds to a linear spring, which is described by the following relation

$$f = -k^* \left( y - y_0^* \right), \tag{2}$$

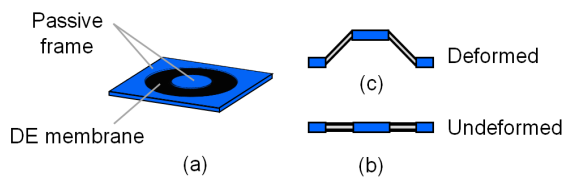

Fig. 1. DE membrane (a), undeformed (b) and deformed (c) states in cross-sectional views. Outer radius, inner radius, and thickness in the undeformed state are equal to 11 mm, 6.25 mm, and 40 μm respectively.

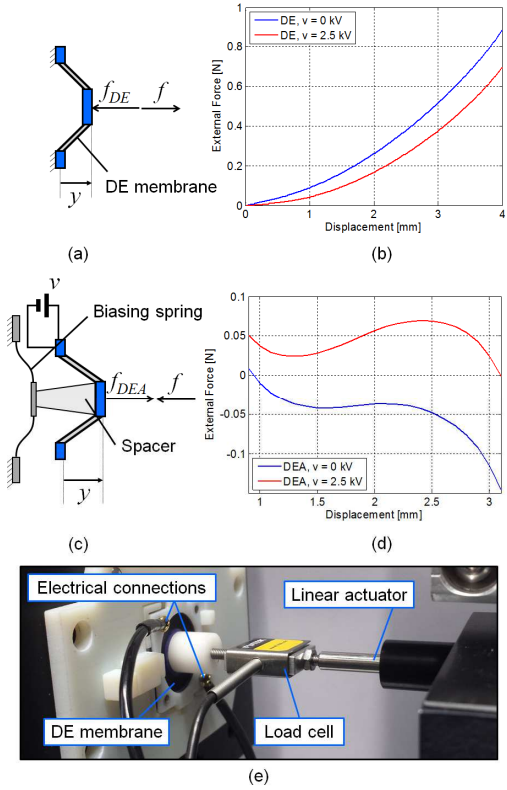

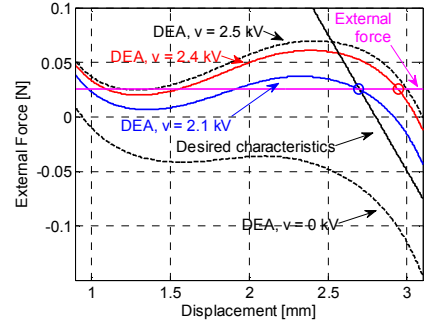

Fig. 3. Explanation of IC principle for the DEA.

may exhibit multiple intersections between external force and DEA curves (equilibrium points). If multiple equilibria exist, it can be shown that the intermediate ones are unstable, making then not possible to control the DEA in open loop.

## III. INTERACTION CONTROL DESIGN

### A. Preliminary notation

The identity matrix is denoted by $\mathbf{I}$, whereas the null matrix is denoted by $\mathbf{0}$. For a matrix $A \in \mathbb{R}^{m \times n}$, $A^T$ denotes its transpose, while $\text{He}\{A\} = A + A^T$. For two symmetric matrices $A$ and $B$, $A > B$ means that $A - B$ is positive definite. In partitioned symmetric matrices, the symbol $\bullet$ stands for symmetric blocks. The matrix $\text{diag}\{A_1, A_2, \ldots, A_n\}$ is the block-diagonal matrix having $A_1, A_2, \ldots, A_n$ as diagonal blocks. For a vector $x \in \mathbb{R}^n$, $||x||$ denotes the Euclidean norm. For a matrix $A \in \mathbb{R}^{m \times n}$, $||A||$ denotes the induced 2 norm.

### B. DEA modeling

The controller design is based on a detailed and accurate nonlinear and time-invariant model of the DE membrane developed in [16]. The model can be summarized as follows

$$\begin{bmatrix} \dot{x} \\ y \end{bmatrix} = \begin{bmatrix} A(y) & B_F & B_u(y) \\ \hline C & 0 & 0 \end{bmatrix} \begin{bmatrix} x \\ f \\ u \end{bmatrix}, \qquad (3)$$

$x = [x_1 \ x_2 \ x_3]^T$ is the state vector ($x_1$ is the actuator displacement, $x_2$ the speed and $x_3$ is an unmeasurable internal strain related to the viscoelastic behavior), $f$ is the external force, $u$ the control input which corresponds to the square of the voltage $v$ applied to the electrodes of the DE, and $y = x_1$ is the system output. The matrices in (3) are defined as follows

$$A(y) = \begin{bmatrix} 0 & 1 & 0 \\ a_{21}(y) & a_{22}(y) & a_{23}(y) \\ a_{31}(y) & 0 & a_{33} \end{bmatrix}, \qquad (4)$$

$$B_f = \begin{bmatrix} 0 & -1/m & 0 \end{bmatrix}^T, \qquad (5)$$

$$B_u = \begin{bmatrix} 0 & b_{21}(y) & 0 \end{bmatrix}^T, \qquad (6)$$

$$C = \begin{bmatrix} 1 & 0 & 0 \end{bmatrix}. \qquad (7)$$

Differently from standard notation for passive mechanical systems, compressive forces are assumed to be positive. The functions appearing in (4), (6) depend on membrane geometry and constitutive material parameters, and exhibit a polynomial where $f$ is the external compressive force, $y$ is the actuator displacement, while the scalars $y_0^*$ and $k^*$ are design variables that allow to parameterize the desired force-displacement interaction profile. The basic principle of the IC can be introduced considering the example shown in Fig. 3, in which the desired linear force-displacement characteristics is depicted in black solid line. Suppose that an external compressive force of 0.025 N is applied to the DE (magenta line) and the membrane is subject to a voltage of 2.4 kV (the corresponding DEA curve is depicted in red). The resulting equilibrium state (red circle) is provided by the intersection between external force and DEA curve. As it can be observed, the equilibrium does not belong to the desired characteristics (i.e., the black line). In order to let the equilibrium lie on the desired characteristics, the voltage has to be reduced to the value of 2.1 kV. This decrease in voltage modifies the DEA characteristics to a new function (in blue), which intersects the external load exactly in correspondence of the desired force-displacement characteristics (blue circle). Thus, from the external force viewpoint, the DEA is mechanically equivalent to the spring described by the black line. If the external force changes, the voltage $v$ needs to be modified accordingly to maintain the resulting equilibrium on the desired characteristics, at least in the steady state. Note that it is always possible to find a voltage which satisfies this condition, as long as the system operates in the area delimited by the DEA characteristics for minimum and maximum voltage (dashed lines in Fig. 3). Moreover, due to the non-monotonic behavior of the DEA characteristics the system dependency on $y$ (see [16] for details). We proved in [16] that if bounds on $u$ are known, the output $y$ is confined in a known bounded interval, in which all functions in (4), (6) are smooth (the proof can be extended straightforwardly to account for $f$ as well). Under these conditions, the quasi-LPV structure of (3) lends itself to a control design based on LMI tools [34].

Fig. 2. Unbiased DE membrane picture (a), force-displacement characteristics for different voltages (b), DEA picture (c), force-displacement characteristics for different voltages (d), and photo of the experimental setup (e).

### C. IC specification

In the general case, IC algorithms should allow the DE to reproduce any kind of desired force-displacement (or force-velocity) characteristics, either linear, nonlinear, or dynamic. Since model (3) admits a causal representation only in case the force is considered as input, it is somehow natural to express the IC specification in terms of a force-input, displacement-output characteristics. We assume that the desired mechanical characteristics can be expressed in the following LPV form

$$\begin{bmatrix} \dot{x}_i \\ y^* \end{bmatrix} = \left[ \begin{array}{c|cc} A_i(p) & B_{f,i}(p) & B_{y_0^*,i}(p) \\ \hline C_i(p) & D_{f,i}(p) & D_{y_0^*,i}(p) \end{array} \right] \begin{bmatrix} x_i \\ f \\ y_0^* \end{bmatrix}, \quad (8)$$

where $p$ represents a measurable, generally time-varying parameter, $x_i \in \mathbb{R}^{n_i}$ is an internal state, $n_i$ is a positive integer to be defined, $y^*$ is the ideal displacement corresponding to the force $f$, and $A_i(p)$, $B_{f,i}(p)$, $B_{y_0^*,i}(p)$, $C_i(p)$, $D_{f,i}(p)$, $D_{y_0^*,i}(p)$ are matrix functions of $p$ of appropriate dimensions. Note that $p$ acts as a scheduling parameter for the LPV model (8), while $y_0^*$ represents an exogenous input, e.g., a displacement set-point in case of zero contact force.

Model (8) can be used to parameterize a large family of linear and nonlinear desired behaviors. Two cases of particular interest are discussed in the sequel. The first one consists of the following static nonlinear force-displacement curve

$$y^* = y_0^* - k^*(y)^{-1} f, \quad (9)$$

where $p = y$, $k(y)^* > 0$, $\forall y$. Note that (9) represents a special case of (8), since no dynamics are involved, while at the same time it is a generalization of (2), since the stiffness $k^*(y)$ is allowed to be a function of the deformation. Another specification of particular interest corresponds to the case in which the actuator reacts to an external force as a linear mass-spring-damper device. In this case, the model results into

$$\begin{bmatrix} \dot{x}_{i,1} \\ \dot{x}_{i,2} \\ y^* \end{bmatrix} = \left[ \begin{array}{cc|cc} 0 & 1 & 0 & 0 \\ -k^*/m^* & -b^*/m^* & 1/m^* & k^*/m^* \\ \hline 1 & 0 & 0 & 0 \end{array} \right] \begin{bmatrix} x_{i,1} \\ x_{i,2} \\ f \\ y_0^* \end{bmatrix}. \quad (10)$$

### D. IC design

The goal of this section is to design a control law $u$ for (3) in such a way that the closed loop system mimics the behavior of (8). We introduce first the interaction error $e_i$

$$e_i = y - y^*, \quad (11)$$

where $y^*$ is specified by (8). Note that if $e_i = 0$, then (8) is verified for $y^* = y$, and therefore the DEA reacts to external force $f$ according to the desired behavior. Thus, the feedback law has to drive $e_i$ as close as possible to zero.

Fig. 4 shows the structure of the proposed controller. It consists of three modules (blue box, Fig. 4). Two of them are placed in cascade with the plant to form an augmented plant (red box, Fig. 4). The first module computes the interaction error as a function of $f$ and $y$, as defined by (8) and (11). The second module is an error shaping filter with state realization

$$\begin{bmatrix} \dot{x}_s \\ z \end{bmatrix} = \left[ \begin{array}{c|c} A_s(p) & B_s(p) \\ \hline C_s(p) & D_s(p) \end{array} \right] \begin{bmatrix} x_s \\ e_i \end{bmatrix}, \quad (12)$$

with state $x_s \in \mathbb{R}^{n_s}$, $n_s$ is a positive integer to be defined, input $e_i$, output $z$, and $A_s(p)$, $B_s(p)$, $C_s(p)$, $D_s(p)$ are matrix functions of appropriate dimensions to be designed. The shaping filter can be used to express the desired closed loop performance as an upper bound on the norm from exogenous inputs $f$ and $y_0^*$ to $z$. By properly tuning the gain of the shaping filter, it is possible to penalize the interaction error at certain frequencies in order to make the controller more accurate within a desired bandwidth, similarly to the standard practice in $\mathcal{H}_\infty$ control [35]. Specifically, the gain of the shaping filter can be made infinitely large at null frequency by selecting one of the eigenvalues of $A_s(p)$ equal to zero, in order to ensure zero error at steady-state even presence of constant disturbances or model uncertainties (integral control).

The overall state realization of the augmented plant can be written as in (13) at the bottom of this page, where the matrices $C_{m1}$, $C_{m2}$, and $C_{m3}$ are selected in such a way that the output $x_m$, i.e., the measurable states, is defined as follows

$$x_m = \begin{bmatrix} x_s & x_i & x_1 & x_2 \end{bmatrix}^T, \quad (14)$$

excluding in this way the unmeasurable state $x_3$. System (13) can be rewritten in a more compact form as

$$\begin{bmatrix} \dot{x}_a \\ z \\ x_m \end{bmatrix} = \left[ \begin{array}{c|ccc} A_a(p,y) & B_{f,a}(p) & B_{y_0^*,a}(p) & B_{u,a}(y) \\ \hline C_{z,a}(p) & D_{f,a}(p) & D_{y_0^*,a}(p) & 0 \\ C_{m,a} & 0 & 0 & 0 \end{array} \right] \begin{bmatrix} x_a \\ f \\ y_0^* \\ u \end{bmatrix}, \quad (15)$$

$$\begin{bmatrix} \dot{x}_s \\ \dot{x}_{,i} \\ \dot{x} \\ \hline z \\ x_m \end{bmatrix} = \left[ \begin{array}{ccc|ccc} A_s(p) & -B_s(p)C_i(p) & B_s(p)C & -B_s(p)D_{f,i}(p) & -B_s(p)D_{y_0^*,i}(p) & 0 \\ 0 & A_i(p) & 0 & B_{f,i}(p) & B_{y_0^*,i}(p) & 0 \\ 0 & 0 & A(y) & B_f & 0 & B_u(y) \\ \hline C_s(p) & D_s(p)C_i(p) & D_s(p)C & -D_s(p)D_{f,i}(p) & -D_s(p)D_{y_0^*,i}(p) & 0 \\ C_{m1} & C_{m2} & C_{m3} & 0 & 0 & 0 \end{array} \right] \begin{bmatrix} x_s \\ x_i \\ x \\ \hline f \\ y_0^* \\ u \end{bmatrix} \quad (13)$$

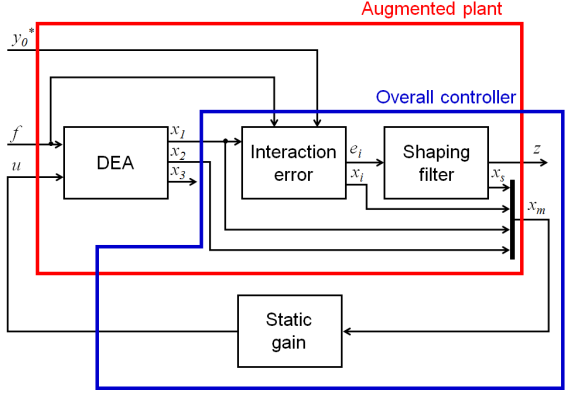

Fig. 4. Block diagram of the closed loop system.

where the augmented state $x_a$ is defined as

$$x_a = \begin{bmatrix} x_s & x_i & x \end{bmatrix}^T, \qquad (16)$$

while all the other matrices appearing in (15) can be obtained by a comparison with (13). Note also that the following partitioning holds

$$x_a = \begin{bmatrix} x_s & x_i & x_1 & x_2 \mid x_3 \end{bmatrix}^T = \begin{bmatrix} x_m \mid x_3 \end{bmatrix}^T. \qquad (17)$$

The third component of the controller is the static partial state feedback law on the measurable state $x_m$,

$$u = -Kx_m, \qquad (18)$$

as depicted in Fig. 4, where $K$ is a constant gain vector to be designed. Note that the proposed controller does not require either the measurement of the full plant state or estimates of the unmeasured components. This renders its implementation far easier. The overall implementation scheme of the resulting dynamic controller encircled by the solid blue line in Fig. 4.

By replacing (18) in (15), the resulting closed loop system is obtained. The gain $K$ has to be designed such that the closed loop system satisfies the following conflicting specifications:

1) Keep the $\mathcal{L}_2$ gain from the exogenous input $w$ to $z$ smaller than a desired upper bound $\lambda > 0$, to minimize the effects of exogenous inputs on filtered interaction error. The exogenous inputs can be related to $f$ and $y_0^*$ in many possible ways, e.g., $w = f$, $w = y_0^*$, $w = [f \ y_0^*]^T$, or eventually $w$ is a weighted combination between $f$ and $y_0^*$;

2) Minimize the weighted Euclidean norm of the controller gain $K$, given by $\|KW\|$, where $W > 0$ is selected by the designer, to reduce control effort and noise amplification.

From now on, it will be assumed for simplicity that $w = f$, since in this work we are primarily interested in controlling the DEA as a "programmable spring". Therefore, we assume that $y_0^*$ is constant while $f$ represents a general exogenous input. Nevertheless, the results presented in the sequel can be extended to more general settings, for which, e.g, $y_0^*$ can be assumed to be time-varying. Note also that the proposed strategy accounts for the stability and dynamic performance of the overall closed loop system by taking into account that $f$ acts at the same time as a disturbance for the DEA and as a command signal for the controller.

The design of a control law ensuring both specifications can be recast into the following BMI optimization problem [34]:

find a scalar $\gamma$, matrix $P > 0$, and a rectangular matrix $K$ of appropriate dimensions such that $\gamma^2$ is minimized and

$$\begin{bmatrix} \text{He}\{A_a(p,y)P - B_{u,a}(p)KC_{m,a}P\} & B_{f,a}(y) & PC_{z,a}(p)^T \\ \bullet & -\mathbf{I} & D_{f,a}(p)^T \\ \bullet & \bullet & -\lambda^2\mathbf{I} \end{bmatrix} < 0, \quad (19)$$

$$\begin{bmatrix} \mathbf{I} & W^T K^T \\ \bullet & \gamma^2 \mathbf{I} \end{bmatrix} > 0, \qquad (20)$$

hold for every admissible $y$, $p$. Due to the nonlinear dependence of system matrices from $y$, the problem is addressed numerically by gridding techniques [36]. We point out that (19) is related to specification 1), while (20) corresponds to 2).

To sum up, the design parameters of the controller are $A_i(p)$, $B_{f,i}(p)$, $B_{y_0^*,i}(p)$, $C_i(p)$, $D_{f,i}(p)$, $D_{y_0^*,i}(p)$, $A_s(p)$, $B_s(p)$, $C_s(p)$, $D_s(p)$, $\lambda$, $W$, and $K$. Scheduling variable $p$ can contain any measurable quantity, e.g., the actuator position, velocity, or external force. At first, matrices $A_i(p)$, $B_{f,i}(p)$, $B_{y_0^*,i}(p)$, $C_i(p)$, $D_{f,i}(p)$, $D_{y_0^*,i}(p)$ are selected to describe the desired DEA mechanical characteristics. Then, tuning parameters $A_s(p)$, $B_s(p)$, $C_s(p)$, $D_s(p)$, $\lambda$, and $W$ are chosen to specify a desired closed loop performance. Finally, the gain $K$ ensuring the desired specification is determined by solving (19)-(20). In some cases, iterations between the last two steps are needed to get a satisfactory transient behavior.

### E. Design problem convexification

Solving the optimization problem (19)-(20) leads to a control law ensuring the desired specifications. However, due to BMI constraint, such a problem is not appealing from a numerical standpoint. To overcome this issue, in this section we come up with a convex relaxation of (19)-(20). Pursuing this approach allows to render the problem numerically tractable, though at the expense of additional conservatism.

BMI (19) corresponding to specification 1) is nonconvex due to the bilinear term in $P$ and $K$ appearing in the (1,1) block. This is a consequence of the nature of the controller, which appears in the form of a partial state feedback law for the augmented plant (15). To linearize such a BMI, we recall the partitioning provided by the right-hand side of (17), where $x_3$ represents the unmeasurable state. By replacing (18) in (15) and partitioning the closed loop system according to (17), we obtain

$$\begin{bmatrix} \dot{x}_m \\ \dot{x}_3 \\ \hline z \end{bmatrix} = \begin{bmatrix} A_{1,a}(p,y) - B_{u,a}(y)K & A_{2,a}(y) & B_{f,a}(p) & B_{y_0^*,a}(p) \\ A_{21,a}(y) & A_{22,a} & \mathbf{0} & \mathbf{0} \\ \hline C_{1z,a}(p) & \mathbf{0} & D_{f,a}(p) & D_{y_0^*,a}(p) \end{bmatrix} \begin{bmatrix} x_m \\ x_3 \\ f \\ y_0^* \end{bmatrix}. \quad (21)$$

As suggested in [37], we consider a block-diagonal positive definite matrix $P$ partitioned according to (21), as follows

$$P = \text{diag}\{P_1, P_2\}, \qquad (22)$$

and a row vector $Y$ given by

$$Y = KP_1. \qquad (23)$$

By replacing (22)-(23) into (19), the following LMI is obtained

$$\begin{bmatrix} H(p,y) & A_{12,a}(y)P_2 + P_1 A_{21,a}(y)^T & B_{f,a}(p) & P_1 C_{1z,a}(p)^T \\ \bullet & A_{22,a}P_2 + P_2 A_{22,a}^T & \mathbf{0} & \mathbf{0} \\ \bullet & \bullet & -\mathbf{I} & D_{f,a}(p)^T \\ \bullet & \bullet & \bullet & -\lambda^2 \mathbf{I} \end{bmatrix} < 0, \quad (24)$$

where

$$H(p,y) = \mathrm{He}\left\{ A_{11,a}(p,y)P_1 - B_{1u,a}(y)Y \right\}. \quad (25)$$

The structural constraint imposed on $P$ by (22) represents the key step that permits to reformulate (19) as (24). Obviously, this constraint on matrix $P$ entails additional conservatism. We also remark that (24) strongly requires that $A_{22,a}$ is Hurwitz. This is always true for DE systems, since this matrix describes the viscoelastic dynamics of the material, which is an inherently stable phenomena.

By solving (23) for $K$ and replacing the resulting value in (20), the resulting inequality becomes

$$\begin{bmatrix} \mathbf{I} & W^T P_1^{-1} Y^T \\ \bullet & \gamma^2 \mathbf{I} \end{bmatrix} > 0, \quad (26)$$

that is nonlinear in the decision variables $P_1$ and $Y$. A possible way to overcome this problem is given by the following result.

*Proposition 1.* Let be $W > 0$. If there exist a symmetric matrix $P_1 > 0$, a real scalar $\gamma > 0$, and a rectangular matrix $Y$ such that

$$\begin{bmatrix} P_1 W^{-T} + W^{-1} P_1 - \mathbf{I} & Y^T \\ \bullet & \gamma^2 \mathbf{I} \end{bmatrix} > 0, \quad (27)$$

then

$$\left\| Y P_1^{-1} W \right\| \equiv \left\| K W \right\| < \gamma. \quad (28)$$

*Proof of Proposition 1.* By pre-and-post multiplying the left-hand side of (26) by $\mathrm{diag}\{P_1 W^{-T}, \mathbf{I}\}$ and $\mathrm{diag}\{W^{-1} P_1, \mathbf{I}\}$ respectively we obtain

$$\begin{bmatrix} P_1 W^{-T} W^{-1} P_1 & Y^T \\ \bullet & \gamma^2 \mathbf{I} \end{bmatrix} > 0. \quad (29)$$

Moreover, given any real square matrix $Q$, the following relation holds

$$(Q - \mathbf{I})^T (Q - \mathbf{I}) \geq 0, \quad (30)$$

thus implying

$$Q^T Q \geq Q^T + Q - \mathbf{I}. \quad (31)$$

Therefore, by setting $Q = W^{-1} P_1$ we have that the satisfaction of (27) implies the satisfaction of (26), thus concluding the proof. $\square$

Note that (27) is linear in all the decision variables, therefore the complete design problem can be addressed by efficient noniterative numerical solvers [34]. On the other hand, notice that (27) introduces some conservatism. Therefore, the minimum value of $\gamma$ for which (27) holds defines only an upper bound on the minimum value of $\gamma$ fulfilling (20).

The design of the controller can then be expressed as the following LMI eigenvalue problem: find a scalar $\gamma$, matrices $P_1 > 0$, $P_2 > 0$, and a rectangular matrix $Y$ of appropriate dimensions such that $\gamma^2$ is minimized and (24), (27) hold for every admissible $y$, $p$. Once the problem is solved, the resulting controller is given by

$$K = Y P_1^{-1}. \quad (32)$$

Despite the LMI design problem can be addressed by efficient solvers, its feasibility is not guaranteed *a priori*. In case the design problem is not feasible, the specification can be made less demanding by tuning the shaping filter. Alternatively, less conservative approaches can be used, such as nonlinear BMI optimization, or different control laws as full state feedback or LPV gain-scheduling. Discussion of these approaches, however, goes beyond the scopes of this paper.

### F. Self-sensing algorithm

The proposed IC law requires both force and displacement feedbacks to be implemented. However, it is possible to reconstruct position $y$ relying on an on-line processing of the available electrical variables. This operation, known as position self-sensing, can be performed in various ways, depending on how the membrane is utilized. In this paper, we use the technique described in [38], summarized as follows.

The electrical response of a DE is typically described by means of an equivalent RC series circuit, i.e.,

$$v = R_e(y) i + \frac{q}{C_e(y)}, \quad (33)$$

where $v$, $q$ and $i$ are, respectively, the DE voltage, charge, and current, with $\dot{q} = i$. Quantities $R_e$ and $C_e$ represent the deformation-dependent equivalent resistance and capacitance of the DE, respectively. The equivalent capacitance manifests a monotonic and non-hysteretic dependence on the membrane displacement, as shown in the experimental curve in Fig. 5. Voltage and current measurements are typically available online, while measuring the charge during actuation is more challenging. If $v$ and $i$ are used to reconstruct the equivalent capacitance $C_e$ during actuation, it is possible to estimate the actuator deformation without an additional electro-mechanical transducer, therefore achieving self-sensing. In principle, one could reconstruct the charge $q$ by integrating $i$, and therefore exploiting the linear-in-parameter structure of equation (33) to estimate $R_e$ and $C_e$ via standard linear regression algorithms, such as Recursive Least Squares (RLS). However, the measurement bias results into an integration drift, making it not possible to obtain $q$ via integration. Therefore, the direct implementation of online estimation techniques based on equation (33) is not possible if charge measurements are not available (such as in our case). On the other hand, low-frequency components of voltage are typically filtered by (33) (since it exhibits a high-pass nature), therefore the measured current arising from a typical low-frequency actuation turns out to be too small to be accurately measured. To address this issue, a low-amplitude, high-frequency sinusoidal voltage component is superimposed to the actuation voltage. Such a high-frequency signal does not produce actuation, as it is filtered by the mechanical bandwidth of the actuator, but at the same time it produces a large current which can be accurately measured. Additionally, if the high-frequency signal is much faster than the actuator motion, electrical quantities in (33) evolve in time much faster than parameters $R_e$ and $C_e$, which depend on mechanical deformation. This property can be exploited for self-sensing. By differentiating in time equation (33), we obtain

$$\dot{v} \cong R_e(y)i + \frac{i}{C_e(y)}, \qquad (34)$$

where time derivatives of $R_e$ and $C_e$ have been neglected, being the variation of such slow parameters much smaller than high-frequency voltage and current variations. Then, RLS estimation can be performed on equation (34) to obtain $R_e$ and $C_e$, since only current and voltage measurements are required. The exponential forgetting of RLS permits the estimation of time-varying $R_e$ and $C_e$ as well, provided that they are significantly slower than the high-frequency self-sensing signal. Additional digital filtering techniques, such as low-pass and comb filters, can be also implemented to mitigate the effects of measurement noise and harmonic disturbances.

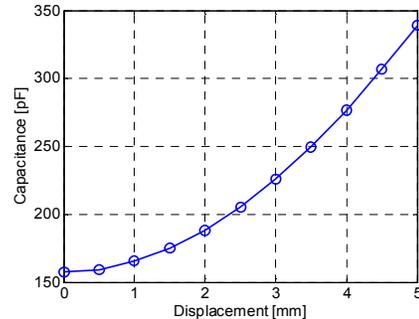

Fig. 5. DEA measured capacitance for different deformations [38].

## IV. EXPERIMENTAL VALIDATION

### A. Experimental setup

The proposed IC architecture is validated on the experimental test rig shown in Fig. 6. A Zaber T-NA08A25 (Linear Actuator 1) is used to control the compression of the bi-stable biasing spring towards the DE membrane, allowing to tune the actuator stroke. A second linear actuator (Linear Actuator 2), namely an Aerotech ANT 25-LA, is used to apply a desired force to the DEA. To control this force, a Futek LSB-200 load cell is attached to the end of the linear actuator, and the motor current is regulated with a PID algorithm such that the measured contact force equals a desired reference signal. A Trek 610E voltage amplifier is used to apply the control voltage to the DEA. An embedded sensor on the amplifier allows to perform voltage measurements, while an external measurement circuit has been designed in order to measure the electric current. Real-time digital acquisition and control algorithms are run on a National Instruments FPGA board programmable in LabVIEW. All the measurements are acquired at a fixed rate of 20 kHz. DEA IC and motor force control algorithms are executed and synchronized in real time with constant rates of 5 kHz and 2 kHz, respectively. Back-calculation anti-windup algorithms are also implemented to deal with saturation limits. The self-sensing is implemented in real-time with sampling frequency of 20 kHz, and synchronized with the control loops.

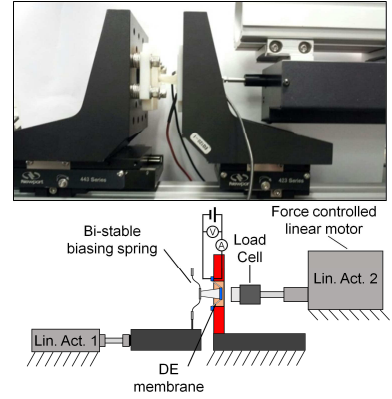

Fig. 6. DEA experimental setup, picture (upper) and sketch (lower).

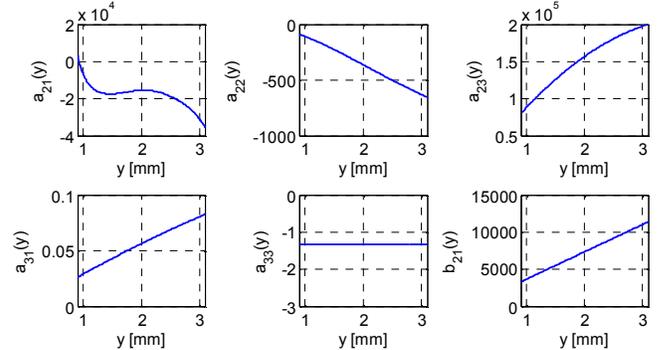

Fig. 7. Identification for nonlinear functions in (4) and (6).

### B. System identification and controller tuning

A preliminary parameter identification procedure is initially performed. The resulting model functions are reported in Fig. 7. Furthermore, the actuator spacer mass $m$ affecting (5) equals 2.05 g. The model is subsequently used to design the IC algorithms according to the approach discussed in Section III. Several control laws are designed, based on different specifications. For each design, desired mechanical response (8) and error shaping filter (12) need to be determined by the designer, while the controller gain appearing in (18) is determined by solution of (24), (27). The definition of the desired mechanical characteristics depends on the particular specification for the given interaction task. Following the design rules reported in [35], the corresponding selections for the shaping filter are reported in Table I. Note that all filters are parameterized as functions of $\omega_k$ and $M_s$, representing tuning parameters that have to be selected in order to find the best trade-off between settling time, oscillations, and control input saturation. Furthermore, values of $\lambda = 1$ and $W = \mathbf{I}$ are selected for all the proposed design.

TABLE I
DEFINITIONS OF SHAPING FILTER FOR EACH SPECIFICATION

| Specification | $A_s(p)$ | $B_s(p)$ | $C_s(p)$ | $D_s(p)$ |
|---|---|---|---|---|
| Linear profile, static equation (9) | 0 | 1 | $k^*\omega_k$ | $k^*/M_s$ |
| Nonlinear profile, static equation (9) | 0 | 1 | $k^*(y)\omega_k$ | $k^*(y)/M_s$ |
| Linear profile, dynamic equation (10) | 0 | 1 | $k^*\omega_k$ | 0 |

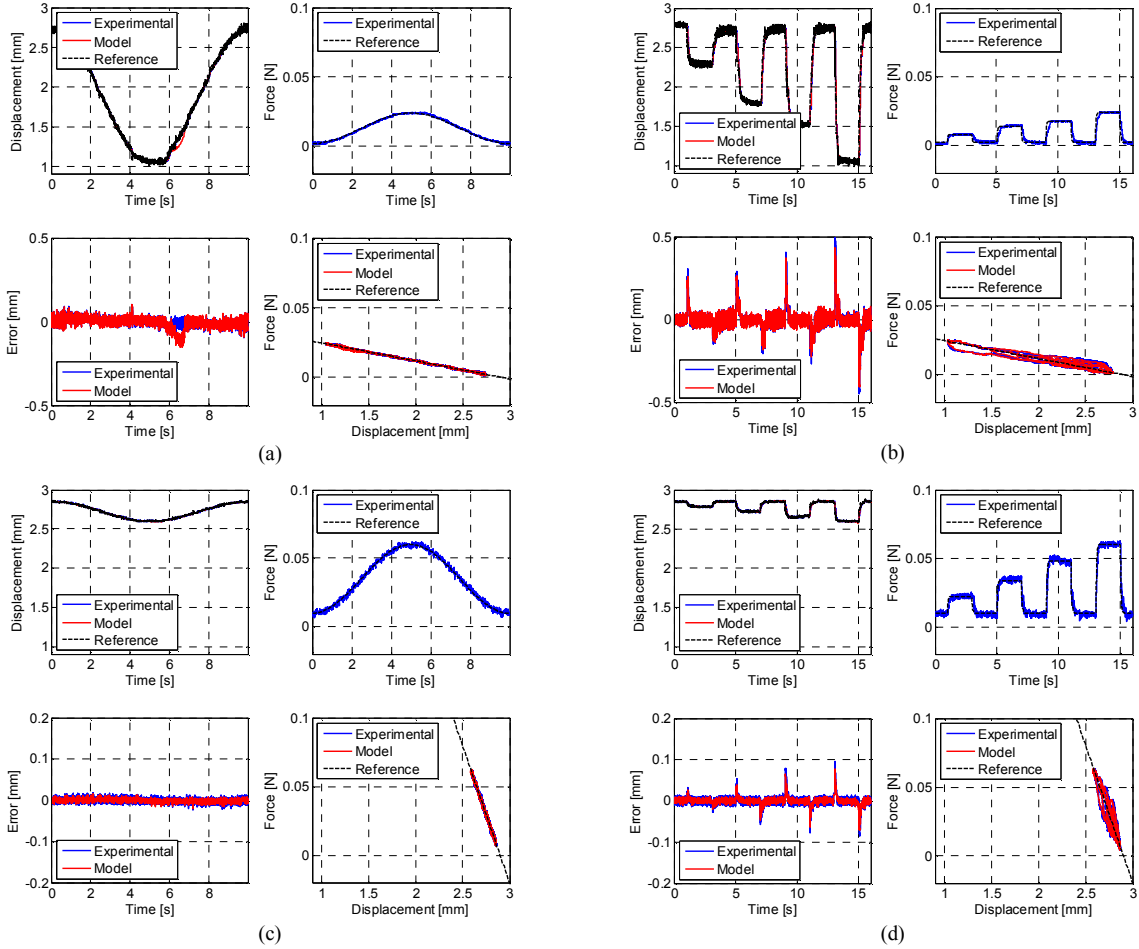

Fig. 8. DEA IC, experimental (blue) and simulation (red) results, linear profile with $k^* = 0.013$ N/mm, $y_0^* = 2.9$ mm, sinusoidal (a) and AM square wave force (b), linear profile with $k^* = 0.20$ N/mm, $y_0^* = 2.9$ mm, sinusoidal (c) and AM square wave force (d).

### C. Results, linear force-displacement profile

The first experiments aim at imposing a linear force-displacement characteristics to the DEA. The experiments consist of applying a desired force by means of the linear actuator, and verifying whether the DEA reacts to it according to the desired mechanical characteristics. The linear profile with a constant $k^* = 0.013$ N/mm and $y_0^* = 2.9$ mm for the first controller, $k^* = 0.2$ N/mm and $y_0^* = 2.9$ mm for the second controller. For both controllers, the shaping filter is designed as reported in Table I, where $A_s(p) = 0$ to introduce integral control, while $\omega_s$ and $M_s$ permit to shape the transient behavior. By means of simulations, satisfactory values of $\omega_s = 15$ and $M_s = 2$ are selected for each design. Simulation and experimental results are reported for $k^* = 0.013$ N/mm in Fig. 8(a) and (b), and for $k^* = 0.2$ N/mm in Fig. 8(c) and (d). Each figure shows the reference and measured force, the resulting displacement, the interaction error, and the force-displacement trajectory. Two different force inputs are tested, namely a 0.1 Hz sinewave and an AM square wave. As confirmed by simulations and experiments, both controllers provide satisfactory results in all of the tests.

In particular, the resulting DEA is able to approximate the desired force-displacement characteristics (black dashed line) with satisfactory accuracy. For the sinewave tests, the system evolves onto the desired curve almost every time. It can be observed that increasing the frequency of the applied force makes the interaction error larger, as expected from a typical low-pass closed loop behavior. Such an error can be reduced by properly modifying the shaping filter, or by adopting more complex nonlinear control laws. This aspect is one of the open issues that will be considered in future research. For the AM square wave, the interaction error converges to zero at steady state after an exponential transient. It is worthwhile to notice that such an exponential convergence is mainly due to the dynamics of the closed loop system rather than to the material creep. In fact, while the material creep makes the position drift for several seconds after the application of the voltage (see [16]), in case of controlled DEA the system reaches the steady values in a significantly faster time, i.e., order of 0.2-0.3 s in Fig. 8. The effects of viscoelasticity, mainly due to the internal unmeasurable state $x_3$, are thus effectively compensated by means of the feedback of the other state variables. As a final remark, it should be noted that, for the controlled DEA with lower stiffness, the controller allows stable operation in the region that is unstable in open loop (where the DEA curve exhibits a positive slope, see Fig. 2(d)), thus overcoming one of the major challenges of the considered actuator.

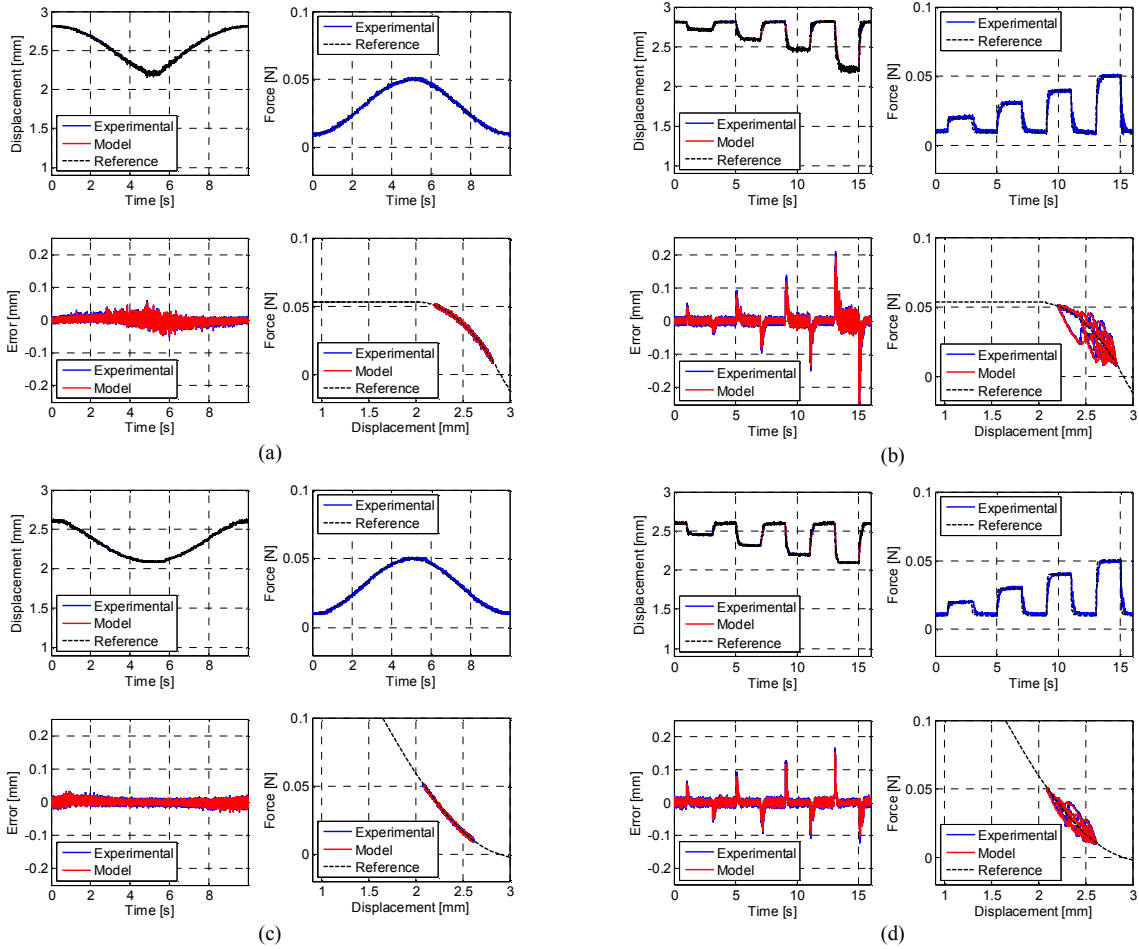

Fig. 9. DEA IC, experimental (blue) and simulation (red) results, nonlinear softening profile, sinusoidal (a) and AM square wave force (b), nonlinear stiffening profile, sinusoidal (c) and AM square wave force (d).

### D. Results, nonlinear force-displacement profile

Experiments with nonlinear force-displacement profiles are presented in this section. Fig. 9(a) and (b) show the response in case of softening force-displacement profile, while Fig. 9(c) and (d) show results for a stiffening profile. In both cases, $y_0^*$ = 2.9 mm, and the tuning is performed with the shaping filter reported in Table I, by selecting $\omega_b$ = 15 and $M_s$ = 2. The controllers are tested by taking as a reference force both a sinusoidal signal and an AM square wave. Thanks to the proposed controller, the DEA is capable of replicating the desired profile with satisfactory accuracy. Note that, in this case, the instantaneous stiffness of the closed loop system is not constant, but it changes continuously according to the operating point. Interestingly, it can be noted that the interaction error increases when the stiffness decreases, and decreases when the stiffness increases.

### E. Results, dynamic force-displacement profile

Finally, Fig. 10 shows results in case a dynamic force-displacement specification is selected, namely a linear mass-spring-damper system, as described by equation (10). The values of $k^*$ and $y_0^*$, describing the static force-displacement response, are selected equal to 0.1 N/mm and 2.9 mm in each test, while the remaining parameters, i.e., $b^*$ and $m^*$, are selected in order to replicate a second order system with given time constant $\tau$ and damping coefficient $\delta$. The same time constant $\tau$ = 2 s is selected in each test, while the damping coefficient is chosen as $\delta$ = 1 in Fig. 10(a), $\delta$ = 0.7 in Fig. 10(b), and $\delta$ = 0.4 in Fig. 10(c). The time constant $\tau$ is chosen compatibly with the dynamics of the force control loop in the linear motor. In this way, the step response of the controlled DEA is at least one order of magnitude slower than the step response of the motor, allowing a better visualization of the IC performance. The shaping filter is selected as in Table I, by setting $\omega_b$ = 1.5. The closed loop behavior is overall satisfactory in each of the considered experiments. However, it can be observed that the displacement exhibits a spike every time the force undergoes a sudden change. This phenomenon, observed in both simulations and experiments, is due to the fact that the proposed control law does not induce a sufficiently fast compensation of the effects of the force on the plant. Clearly, these spikes can be reduced by properly tuning the controller. However, it is observed experimentally that if the closed loop system is made excessively fast, the performance start to degrade due to the delay introduced by the self-sensing (results are omitted for conciseness).

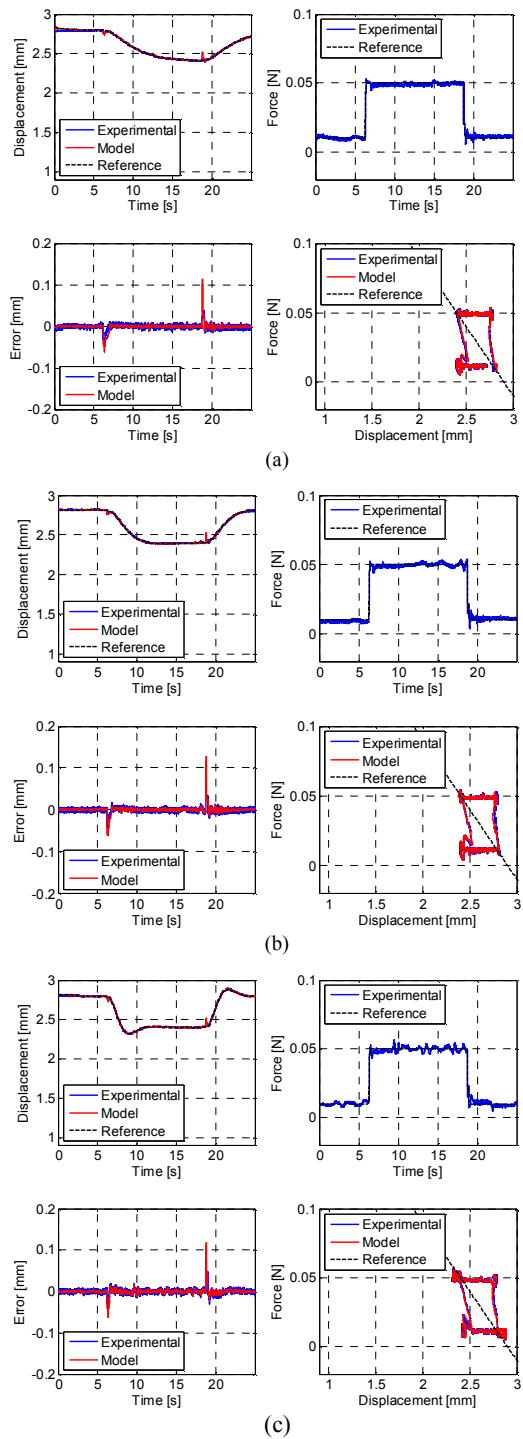

Fig. 10. DEA IC, experimental (blue) and simulation (red) results, step force, dynamic force-displacement profile, $\tau = 2$ s and $\delta = 1$ (a), $\tau = 2$ s and $\delta = 0.7$ (b), $\tau = 2$ s and $\delta = 0.4$ (c).

Therefore, both of these aspects must be taken into account during the design.

## V. Conclusion

This paper has presented the design of a robust interaction control law for a bi-stable dielectric elastomer membrane actuator. The proposed algorithm allows to control the voltage in a way such that the elastomer reacts to an external force similarly to a mechanical system described by a desired force-displacement profile. The control law has been implemented in conjunction with a self-sensing algorithm that allows to achieve the desired specification by requiring contact force measurements only, reducing the number of electromechanical transducers employed in the system. One appealing feature of our solution is that the proposed controller requires a relatively low implementation effort and provides an optimal trade-off between performance and control effort minimization. Our experiments revealed that the performance shows some initial degradation only in case a dynamic behavior is assigned. This may represent a limitation of the proposed linear strategy, but it can be overcome by further extensions of our approach based on more advanced strategies, e.g., nonlinear methods. Moreover, we plan to work on the development of more advanced estimation techniques implementing both force and position self-sensing for a fully sensorless interaction control approach.

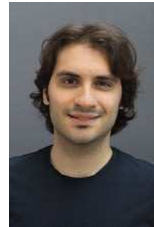

**Gianluca Rizzello** (M'16) was born in Taranto, Italy, in 1987. He received the master's (Hons.) degree in control engineering from the Polytechnic of Bari, Italy, in 2012. He received his Ph.D. in Information and Communication Technologies from Scuola Interpolitecnica di Dottorato, a joint program between Polytechnic Universities of Torino, Bari, and Milano, Italy, in 2016. During his Ph.D. studies he has been a Visiting Research Scholar with the University of Saarland, Saarbrücken, Germany, where he is currently working as a post-doc researcher. His research interests involve modeling, control, and self-sensing of innovative actuators based on smart materials, and control of high-speed electrical motors and generators for aerospace applications.

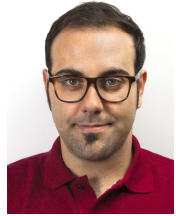

**Francesco Ferrante** (M'16) is a postdoctoral researcher at the Department of Computer Engineering, University of California Santa Cruz. In 2015 and 2016, he held a postdoctoral position at the Department of Electrical and Computer Engineering, Clemson University, South Carolina. He received in 2010 a "Laurea degree" (BS) in Control Engineering from Sapienza University, Rome, Italy and in 2012 a "Laurea Magistrale" degree cum laude (MS) in Control Engineering from University Tor Vergata, Rome, Italy. During 2014, he held a visiting scholar position at the Department of Computer Engineering, University of California at Santa Cruz. In 2015, he received a PhD degree in Control Theory from Institut Supérieur de l'Aéronautique et de l'Espace (SUPAERO) Toulouse, France. His research interests are in the general area of systems and control with a special focus on hybrid systems, observer design, and application of convex optimization in systems and control.

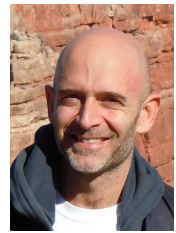

**David Naso** (SM'14) was born in Salerno, on April 29th, 1967. He received the Laurea degree with highest honors in electronic engineering and the Ph.D. degree in electrical engineering, both from the Polytechnic of Bari, Italy, in 1994 and 1998, respectively. During his Ph.D. studies, he has been also a Guest Researcher with the Operation Research Institute, Technical University of Aachen, Germany, in 1997. Since 1999, he has been with Department of Electric and Information Engineering of the Polytechnic of Bari, where he currently serves as Associate Professor and Head of the Robotics lab. His current research interests focus on control of high-speed electrical motors and generators, control of innovative actuators based on smart materials, distributed automation and multi-agent systems.

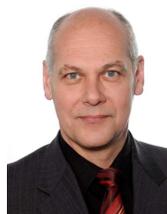

**Stefan Seelecke** received his Ph.D. degree in Engineering Science from Technical University Berlin, Berlin, Germany in 1995. After his habilitation in 1999, he joined the Department of Mechanical and Aerospace Engineering at North Carolina State University, Raleigh, USA, in 2001. He is currently a Full Professor of Systems Engineering and Materials Science & Engineering at Saarland University, Saarbrücken, Germany, where he directs the Intelligent Material Systems Lab. His research interests include the development of smart materials-based actuator and sensor systems, in particular (magnetic) shape memory alloys, piezoelectrics and electroactive polymers.